# Analytical Solution to the Transient Beam Loading Effects of the Superconducting Cavity[*]

Ran Huang(黄燃)[1,2,3;1], Yuan He(何源)[1;2], Zhi-Jun Wang(王志军)[1], Wei-Ming Yue(岳伟明)[1], An-Dong Wu(吴安东)[1,2], Yue Tao(陶玥)[1,2], Qiong Yang(杨琼)[1], Cong Zhang(张聪)[1], Hong-Wei Zhao(赵红卫)[1], Zhi-Hui Li(李智慧)[3]

[1] Institute of Modern Physics, Chinese Academy of Sciences, Lanzhou 730000, China

[2] University of Chinese Academy of Sciences, Beijing 100049, China

[3] Sichuan University, Chengdu 610064, China

**Abstract:** Transient beam loading effect is one of the key issues in any superconducting accelerators, which needs to be carefully investigated. The core problem in the analysis is to obtain the time evolution of cavity voltage under the transient beam loading. To simplify the problem, the second order ordinary differential equation describing the behavior of the cavity voltage is intuitively simplified to a first order one, with the aid of the two critical approximations lacking the proof for their validity. In this paper, the validity is examined mathematically in some specific cases, resulting in a criterion for the simplification. It's popular to solve the approximated equation for the cavity voltage numerically, while this paper shows that it can also be solved analytically under the step function approximation for the driven term. With the analytical solution to the cavity voltage, the transient reflected power from the cavity and the energy gain of the central particle in the bunch can also be calculated analytically. The validity of the step function approximation for the driven term is examined by direct evaluations. After that, the analytical results are compared with the numerical ones.

# 1. Introduction

For superconducting cavities with heavy beam loading, such as cavities adopted in the prototype Injector II of the C-ADS[1-3], the transient effects resulting from the fierce beam-cavity interaction needs to be carefully investigated in order to provide critical information to the low level control system and the machine protection system to ensure the accelerator can operate expectedly on the normal conditions and avoid accidents when the abnormity occurs. For the past half a century, the equivalent circuit method[4] has been constantly proved to be an effective and precise method in the beam loading analysis, with which a second order ordinary linear differential equation for the cavity voltage can be obtained. For the superconducting cavities, this equation can be further simplified to a first order one. But the two critical approximations used in

[*] Supported by National Natural Science Foundation of China (11525523,91426303)

[1)] E-mail: burn4028@impcas.ac.cn

[2)] E-mail: hey@impcas.ac.cn

the simplification process is somehow intuitive and lack of strictly mathematical proof for its validity. While in this paper, the validity of these two approximations will be examined mathematically in some specific cases. For the simplified equation, it is popular to solve it with numerical methods, while this paper demonstrates that it can also be analytically solved with step function approximation for the driven term, whose validity is examined by the evaluation with the design parameters of the half wave resonator (HWR) adopted in the prototype Injector II of the C-ADS project. With the analytical solution to the cavity voltage, the transient reflected power from the cavity and energy gain of the central particle in the bunch can also be obtained analytically.

Transient beam loading of the half wave resonators (HWR)[5] in the prototype Injector II of the C-ADS project under various situations had been evaluated based on the analytical methods described in this paper, providing critical information to the low level control system and the machine protection system to ensure the accelerator can operate expectedly on the normal conditions and avoid accidents when the abnormity occurs.

## 2. Analytical Analysis

### 2.1 Validity for the Two Approximations

Although the first order ordinary linear differential equation describing the time evolution of the superconducting cavity voltage shows itself quite often on various journal papers and talk slides, the two critical approximations used in the simplification to obtain it from the original second order ordinary linear differential equation is somehow a little intuitive and somewhat lack of strictly mathematical proof for its validity. In this section, the validity of these two approximations will be examined.

The detuning of the cavity $\Delta\omega = \omega - \omega_0$ and loaded half bandwidth of the cavity $\omega_{0.5}$ are always much smaller than the resonant frequency $\omega$ ($\Delta\omega, \omega_{0.5} \ll \omega$) [6], then the following approximation can hold very well,

$$\begin{cases} i\omega + \omega_{0.5} \approx i\omega \\ 2i\omega\omega_{0.5} + 2\omega\Delta\omega + \Delta\omega^2 \approx 2i\omega\omega_{0.5} + 2\omega\Delta\omega \end{cases} \quad (2.1.1)$$

From the equivalent circuit analysis, with (2.1.1) and the relation $\omega_{0.5} \triangleq \omega_0/2Q_L$, the dynamic equation for the phasor of the effective cavity voltage $\vec{V}_a(t)$ can be expressed as[7],

$$\frac{1}{2\omega}\frac{d^2}{dt^2}\vec{V}_a(t) + i\frac{d}{dt}\vec{V}_a(t) + (i\omega_{0.5} + \Delta\omega)\vec{V}_a(t)$$
$$= \frac{\omega_{0.5} r_L}{2}\left[i\vec{I}_c(t) + \frac{1}{\omega}\frac{d}{dt}\vec{I}_c(t)\right] \quad (2.1.2)$$

where $r_L$ are the effective loaded shunt impedance of the cavity, $\vec{I}_c(t)$ is the phasor of the total effective driven current for the cavity voltage.

For a cavity with beam, $\vec{I}_c(t)$ consists of two parts, corresponding to the contribution from the generator and the beam respectively[4],

$$\vec{I}_c(t) = \vec{I}_g(t) + \vec{I}_b(t) \quad (2.1.3)$$

In the beam loading analysis based on the equivalent circuit, it is conventional to lay the phasor of the effective beam current along the positive real axis. Since it is the image current on the cavity wall, not the beam itself, that excites the fields in the cavity, the phasor of the effective driven current from the beam $\vec{I}_b(t)$ should be always along the negative real axis. For a beam with the bunch of longitudinal Gaussian charge distribution in each neighboring bucket, $\vec{I}_b(t)$ can be expressed as[4],

$$\vec{I}_b(t) = -2I_{b0}(t)e^{-\frac{\phi_{0.5}^2}{2}} \quad (2.1.4)$$

where $I_{b0}(t)$ is the DC beam current and $\phi_{0.5} = \omega\sigma_t$ is the half phase width of the bunch (the effect of the beam current intensity on the bunch length is neglected).

And the phasor of the effective driven current from the generator $\vec{I}_g(t)$ can be related to the forward generator power $P_g(t)$ as[4],

$$\vec{I}_g(t) = 4\sqrt{\frac{\beta P_g(t)}{r_c}} e^{i\theta} \quad (2.1.5)$$

where $\beta$ and $r_c$ are the coupling factor and the shunt impedance of the cavity respectively, $\theta$ is the phase of the effective driven current from the generator.

For the RF accelerator, since $\omega$ is usually no less than $10^9$ Hz, it is customary to take the following approximation to simplified (2.1.2),

$$\begin{cases} \vec{I}_c(t) \gg \dfrac{1}{\omega}\dfrac{d}{dt}\vec{I}_c(t) \\ \dfrac{d}{dt}\vec{V}_a(t) \gg \dfrac{1}{2\omega}\dfrac{d^2}{dt^2}\vec{V}_a(t) \end{cases} \quad (2.1.6)$$

Obviously, this approximation is somehow a little intuitive and somewhat lack of strictly mathematical proof for its validity. Therefore, we'll examine the validity of (2.1.6) in some specific cases.

Differentiating (2.1.5) with time to obtain,

$$\frac{d}{dt}\vec{I}_g(t) = 2\sqrt{\frac{\beta}{r_c}} e^{i\theta} \frac{1}{\sqrt{P_g(t)}} \frac{d}{dt} P_g(t) \quad (2.1.7)$$

With (2.1.5) and (2.1.7), we'll have,

$$\frac{\vec{I}_g(t)}{\dfrac{1}{\omega}\dfrac{d}{dt}\vec{I}_g(t)} = \frac{2\omega P_g(t)}{\dfrac{d}{dt}P_g(t)} \quad (2.1.8)$$

Supposing a specific case where the $P_g(t)$ varies linearly with time from zero to $P_g(t)$, then,

$$\frac{d}{dt}P_g(t) = \frac{P_g(t)}{\Delta t} \quad (2.1.9)$$

Substituting (2.1.9) into (2.1.8) to obtain,

$$\frac{\overrightarrow{I_g}(t)}{\frac{1}{\omega}\frac{d}{dt}\overrightarrow{I_g}(t)} = 2\omega\Delta t \tag{2.1.10}$$

The rising time $\Delta t$ of the generator power is typical in the order of $1\mu s$ and $\omega$ is in the order of $1$ GHz, then $2\omega\Delta t$ is on the order of $10^3$, therefore,

$$\frac{\overrightarrow{I_g}(t)}{\frac{1}{\omega}\frac{d}{dt}\overrightarrow{I_g}(t)} \approx 10^3 \gg 1 \tag{2.1.11}$$

or,

$$\overrightarrow{I_g}(t) \gg \frac{1}{\omega}\frac{d}{dt}\overrightarrow{I_g}(t) \tag{2.1.12}$$

For the effective DC current of the beam $I_{b0}(t)$, due to the beam chopping mechanism, it can always be well approximated to a step function, therefore,

$$\frac{d}{dt}I_{b0}(t) \approx 0 \tag{2.1.13}$$

Differentiating (2.1.4) with time to obtain,

$$\frac{d}{dt}\overrightarrow{I_b}(t) = -2e^{-\frac{\phi_{0.5}^2}{2}}\frac{d}{dt}I_{b0}(t) \approx 0 \tag{2.1.14}$$

Therefore,

$$\overrightarrow{I_b}(t) \gg \frac{1}{\omega}\frac{d}{dt}\overrightarrow{I_b}(t) \tag{2.1.15}$$

Adding (2.1.12) and (2.1.15) to obtain,

$$\overrightarrow{I_g}(t) + \overrightarrow{I_b}(t) \gg \frac{1}{\omega}\frac{d}{dt}\left[\overrightarrow{I_g}(t) + \overrightarrow{I_b}(t)\right] \tag{2.1.16}$$

With (2.1.3), the above expression can be written as,

$$\overrightarrow{I_c}(t) \gg \frac{1}{\omega}\frac{d}{dt}\overrightarrow{I_c}(t) \tag{2.1.17}$$

Therefore, the first expression in (2.1.6) has been verified in this specific case.

When the forward generator power, beam current and cavity detuning are fixed, the solution to (2.1.2) under initial condition $\overrightarrow{V_a}(0) = 0$ can be expressed as[8],

$$\overrightarrow{V_a}(t) = \overrightarrow{V_{a,s}}\left(1 - e^{-\frac{t}{\tau}}\right) \tag{2.1.18}$$

Differentiating (2.1.18) with time to obtain,

$$\frac{d}{dt}\overrightarrow{V_a}(t) = \frac{1}{\tau}\overrightarrow{V_{a,s}}e^{-\frac{t}{\tau}} \tag{2.1.19}$$

and,

$$\frac{d^2}{dt^2}\overrightarrow{V_a}(t) = -\frac{1}{\tau^2}\overrightarrow{V_{a,s}}e^{-\frac{t}{\tau}} \tag{2.1.20}$$

With (2.1.19) and (2.1.20), we'll have,

$$\frac{\frac{d}{dt}\vec{V}_a(t)}{\frac{1}{2\omega}\frac{d^2}{dt^2}\vec{V}_a(t)} = 2\omega\tau \tag{2.1.21}$$

The cavity time constant $\tau$ of the superconducting cavity is typically on the order of $1$ ms and $\omega$ is on the order of $1$ GHz, then $2\omega\tau$ is on the order of $10^6$, then,

$$\frac{\frac{d}{dt}\vec{V}_a(t)}{\frac{1}{2\omega}\frac{d^2}{dt^2}\vec{V}_a(t)} \approx 10^6 \gg 1 \tag{2.1.22}$$

or,

$$\frac{d}{dt}\vec{V}_a(t) \gg \frac{1}{2\omega}\frac{d^2}{dt^2}\vec{V}_a(t) \tag{2.1.23}$$

Therefore, the second expression in (2.1.6) has been verified in this specific case.

With (2.1.6), (2.1.2) can be simplified to,

$$\frac{d}{dt}\vec{V}_a(t) + (\omega_{0.5} - i\Delta\omega)\vec{V}_a(t) = \frac{\omega_{0.5} r_L}{2}\vec{I}_c(t) \tag{2.1.24}$$

which is just the equation comprehensively used nowadays in the transient beam loading analysis for the superconducting cavity.

Moreover, with (2.1.10) and (2.1.21), a criterion for simplifying (2.1.2) into (2.1.24) can be summarized as,

$$\begin{cases} 2\omega\Delta t \gg 1 \\ 2\omega\tau \gg 1 \end{cases} \tag{2.1.25}$$

or,

$$\begin{cases} \Delta t \gg \dfrac{T}{4\pi} \\ \tau \gg \dfrac{T}{4\pi} \end{cases} \tag{2.1.26}$$

where $T = 1/f$ is the RF period.

For the case of the normal conducting cavity, the time evolution of its cavity voltage can also be described by (2.1.24), if (2.1.26) can hold well

## 2.2 Cavity Voltage

Since (2.1.24) is a differential equation in the complex domain, it's always preferred to convert it to its equivalence in the real domain for numerical solving purpose, which consists of two equations, corresponding to the real and imaginary part of (2.1.24) respectively. It can be shown such conversion will significantly accelerate the step-size control numerical solving process, such as the widely-used Fehlberg fourth-fifth order Runge-Kutta method[9], especially in the case of extreme detuning. But on the other hand, the conversion will cause unnecessary complexity in mathematical form of the analytical solution. Therefore, (2.1.24) will be directly solved for the analytical solving.

According to the ordinary differential equation theory, the existence of the analytical solution to (2.1.24) mainly depends on the mathematical form of the inhomogeneous term $\vec{I}_c(t)$. It can be proved that (2.1.24) can always be analytically solved via Laplace transformation when $\vec{I}_c(t)$ is a step function, which has the form as the following[10],

$$\vec{I}_c(t) = \begin{cases} I_{c,1} e^{i\vartheta_1} & 0 \leq t < T_1 \\ I_{c,2} e^{i\vartheta_2} & T_1 \leq t < T_2 \\ \dots \\ I_{c,n} e^{i\vartheta_n} & T_{n-1} \leq t < T_n \end{cases} \quad (2.2.1)$$

where $I_{c,n}$ is the amplitude and $\vartheta_n$ is the phase of the effective driven current.

For the superconducting cavity, due to its large cavity time constant[6], $\vec{I}_c(t)$ can always be well approximated into (2.2.1), whose validity is examined in the next section.

As a demonstration here, we'll just write out the analytical solution to (2.1.24) with $\vec{I}_c(t)$ as below,

$$\vec{I}_c(t) = \begin{cases} I_{c,1} e^{i\vartheta_1} & 0 \leq t < T \\ I_{c,2} e^{i\vartheta_2} & t \geq T \end{cases} \quad (2.2.2)$$

With the initial condition $\vec{V}(0) = 0$, the analytical solution can be readily obtained as,

$$\vec{V}_a(t) = \frac{\omega_{0.5} r_L}{2(\omega_{0.5} - i\Delta\omega)} \begin{cases} \vec{A}(t) & 0 \leq t \leq T \\ \vec{B}(t) & t \geq T \end{cases} \quad (2.2.3)$$

where,

$$\vec{A}(t) = \left[1 - e^{-(\omega_{0.5} - i\Delta\omega)t}\right] I_{c,1} e^{i\vartheta_1} \quad (2.2.4)$$

and,

$$\vec{B}(t) = \left[1 - e^{-(\omega_{0.5} - i\Delta\omega)(t-T)}\right] I_{c,2} e^{i\vartheta_2} - \left[1 - e^{(\omega_{0.5} - i\Delta\omega)T}\right] I_{c,1} e^{-(\omega_{0.5} - i\Delta\omega)t + i\vartheta_1} \quad (2.2.5)$$

Solution (2.2.3)-(2.2.5) can be easily verified by substituting them back into (2.1.24).

To facilitate the evaluation, $I_{c,j}$ and $\vartheta_j$ ($j = 1, 2$) in (2.2.5) should be related to the RF and beam parameters.

Substituting (2.1.4) and (2.1.5) into (2.1.3),

$$\vec{I}_c(t) = \vec{I}_g(t) + \vec{I}_b(t) = 4\sqrt{\frac{\beta P_g(t)}{r_c}} e^{i\theta(t)} - 2I_{b0}(t) e^{-\frac{\phi_{0.5}^2}{2}} \quad (2.2.6)$$

With (2.2.6), $I_{c,j}$ and $\vartheta_j$ can be expressed as,

$$I_{c,j} = \left|\vec{I}_{c,j}\right| = 2\sqrt{\left(2\sqrt{\frac{\beta P_{g,j}}{r_c}} \cos\theta_j - I_{b0,j} e^{-\frac{\phi_{0.5}^2}{2}}\right)^2 + \frac{4\beta P_{g,j}}{r_c} \sin^2\theta_j} \quad (2.2.7)$$

$$\vartheta_{\mathrm{j}} = \arg\left(\overrightarrow{I_{\mathrm{c,j}}}\right) = \arctan\left[\left(\cos\theta_{\mathrm{j}} - \sqrt{\frac{r_{\mathrm{c}}}{4\beta P_{\mathrm{g,j}}}}I_{\mathrm{b0,j}}\mathrm{e}^{-\frac{\phi_{0.5}^2}{2}}\right)^{-1}\sin\theta_{\mathrm{j}}\right] \qquad (2.2.8)$$

where $j = 1,\ 2$.

Therefore, with (2.2.3)-(2.2.5), (2.2.7) and (2.2.8), the time evolution of the superconducting cavity voltage under transient beam loading can be solely determined, if the relevant parameters are known.

## 2.3 Transient Reflected Power

One of the necessities to investigate the transient beam loading effects of the superconducting cavity arising from the evaluation for the transient reflected power from the cavity. Due to the drastic interaction between the high intensity beam and the low surface resistance cavity, the peak of the reflected power sometimes can reach several times larger than the forward generator power, possibly causing field emission in the waveguide or overheat in the matched impedance of the circulator. Therefore, the calculation for the reflected power from the cavity under transient beam loading is one of the critical problems in the accelerator engineering. It is of necessity to determine the reflected power under various conditions, in order to take proper measures to avoid such incidents.

Equations for the reflected power from the cavity in various forms appearing on monographs[6, 8], dissertations and journal papers are usually dealing with the steady or quasi-steady states, because they are obtained from the stationary solution to (2.1.2). The quasi-steady state here means the variation of the generator power and the beam current is sufficiently small in the time period comparable to the cavity time constant. For the superconducting cavity, due to its large cavity time constant (usually in the order of 1 ms), the quasi-steady state condition can hardly be satisfied. In another word, for the case of the superconducting cavity, the reflect power $P_{\mathrm{r}} = P_{\mathrm{r}}(t)$ can't be obtained by simply substituting the generator power $P_{\mathrm{g}} = P_{\mathrm{g}}(t)$ and the beam current $I_{\mathrm{b}} = I_{\mathrm{b}}(t)$ into these equations.

There are several ways to calculate the transient reflected power. Among them, method based on the law of the conservation of the energy is the most concise and with the least approximation. For the system consisting of the cavity and the beam, the following relation can hold for any instant under the conservation of the energy[7],

$$P_{\mathrm{g}}(t) = P_{\mathrm{c}}(t) + P_{\mathrm{b}}(t) + P_{\mathrm{r}}(t) + \frac{dU_{\mathrm{c}}(t)}{dt} \qquad (2.3.1)$$

where $P_{\mathrm{c}}(t)$ is cavity dissipation, $P_{\mathrm{b}}(t)$ is the beam power, $P_{\mathrm{r}}(t)$ is the reflected power and $U_{\mathrm{c}}(t)$ is the cavity stored energy.

Then the reflected power $P_{\mathrm{r}}(t)$ can be expressed as,

$$P_{\mathrm{r}}(t) = P_{\mathrm{g}}(t) - P_{\mathrm{c}}(t) - P_{\mathrm{b}}(t) - \frac{dU_{\mathrm{c}}(t)}{dt} \qquad (2.3.2)$$

Substituting the expressions for $P_{\mathrm{c}}(t)$, $P_{\mathrm{b}}(t)$ and $U_{\mathrm{c}}(t)$ in terms of $V_{\mathrm{a}}(t)$ into (2.3.2) to obtain,

$$P_r(t) = P_g(t) - \frac{V_a^2(t)}{r_c} - I_b(t)V_a(t)\cos[\phi(t)] - \frac{2V_a(t)}{\omega\left(\frac{r}{Q}\right)}\frac{dV_a(t)}{dt} \quad (2.3.3)$$

where $T_{tf}$ is the transit time factor.

In (2.3.3), note that $\phi(t) = \arg[\overrightarrow{V_a}(t)]$, then the calculation of the reflected power has been converted into the calculation of the cavity voltage $\overrightarrow{V_a}(t)$. Since $\overrightarrow{V_a}(t)$ is analytically obtained in the previous section, the reflected power $P_r(t)$ can be analytically calculated.

## 2.4 Energy Gain Variation due to the Beam loading

The energy gain $\Delta E_k(t)$ of the central particle in a bunch after passing through the cavity can be calculated as[4],

$$\Delta E_k(t) = q\text{Re}\{\overrightarrow{V_a}(t)\} \quad (2.4.1)$$

where $q$ is the charge of the particle, $T_{tf}$ is the transit time factor, $\text{Re}\{\overrightarrow{C}\}$ means taking the real part of $\overrightarrow{C}$.

$\overrightarrow{V_a}(t)$ will vary due to the beam loading effect, resulting in a varied $\Delta E_k(t)$, according to (2.4.1). With $\overrightarrow{V_a}(t)$ analytically obtained in the previous section, $\Delta E_k(t)$ can also be analytically calculated.

## 3. Validity for the Step Functional Approximation

Note that $\overrightarrow{I_c}(t) = \overrightarrow{I_g}(t) + \overrightarrow{I_b}(t)$. For $\overrightarrow{I_b}(t)$, due to the beam chopping mechanism, it can always be well approximated into a step function as the followed with no doubt.

$$\overrightarrow{I_b}(t) = \begin{cases} I_{b1} & 0 \leq t < t_{b1} \\ I_{b2} & t_{b1} \leq t < t_{b2} \\ I_{b3} & t_{b2} \leq t < t_{b3} \end{cases} \quad (3.1.1)$$

As for $\overrightarrow{I_g}(t)$, the validity for the step approximation can't be self-proven. One straightforward way to check the validity is to compare the results obtained from the continuous function $\overrightarrow{I_{g,c}}(t)$ and its step function approximation $\overrightarrow{I_{g,s}}(t)$.

For $\overrightarrow{I_{g,c}}(t)$, it can take the following smooth form,

$$\overrightarrow{I_{g,c}}(t) = \begin{cases} I_{g,1}e^{i\theta} & 0 \leq t < t_1 \\ I_{g,1} + \left[\dfrac{3(t-t_1)^2}{\Delta t_1^{\,2}} - \dfrac{2(t-t_1)^3}{\Delta t_1^{\,3}}\right](I_{g,2} - I_{g,1})e^{i\theta} & t_1 \leq t < t_1 + \Delta t_1 \\ I_{g,2}e^{i\theta} & t_1 + \Delta t_1 \leq t < t_2 \\ I_{g,2} + \left[\dfrac{3(t-t_2)^2}{\Delta t_2^{\,2}} - \dfrac{2(t-t_2)^3}{\Delta t_2^{\,3}}\right](I_{g,3} - I_{g,2})e^{i\theta} & t_2 \leq t < t_2 + \Delta t_2 \\ I_{g,3}e^{i\theta} & t \geq t_2 + \Delta T \end{cases} \quad (3.1.2)$$

It can be proved that (3.1.2) has a continuous first derivative at each boundary point.

By casting away all the continuous transition parts in (3.1.2), its step function approximation $\overrightarrow{I_{g,s}}(t)$ can be readily obtained as,

$$\overrightarrow{I_{g,s}}(t) = \begin{cases} I_{g,1}e^{i\theta} & 0 \leq t < t_1 \\ I_{g,2}e^{i\theta} & t_1 \leq t < t_2 \\ I_{g,3}e^{i\theta} & t_2 \leq t < t_3 \end{cases} \quad (3.1.3)$$

which takes the form of (2.2.1).

The amplitude of $\overrightarrow{I_{g,c}}(t)$ and $\overrightarrow{I_{g,s}}(t)$ are schematically plotted as the following, for the case where $I_{g,1} > I_{g,3} > I_{g,2}$.

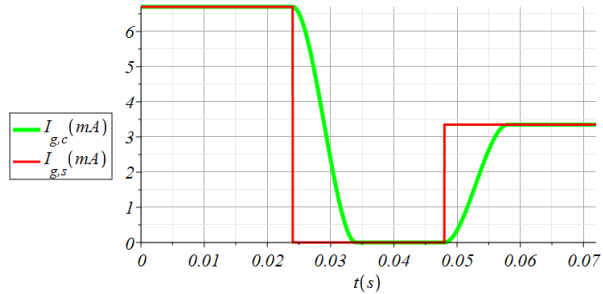

Fig. 1 Schematic plotting of $\overrightarrow{I_{g,c}}(t)$ vs $\overrightarrow{I_{g,s}}(t)$

By substituting (3.1.1) with (3.1.2) or (3.1.3) into (2.1.24) respectively, the resultant equation in both cases can be analytically solved via Laplace transformation.

The rising and dropping edge of $\overrightarrow{I_g}(t)$ are usually less than $40\mu s$, therefore, $\Delta T_1$ and $\Delta T_2$ in (3.1.2) can be set to be $\Delta T_1 = \Delta T_2 = 50us$ for the purpose of comparison.

The validity of the step functional approximation for $\overrightarrow{I_g}(t)$ was examined by evaluation with the design parameters of the half wave resonators (HWR)[5] adopted in the prototype Injector II of the C-ADS project, which are summarized as below,

Table 1 RF and Beam Parameters for the evaluation

| Parameters | Value |
| --- | --- |
| mode frequency $f_0$/MHz | 162.5 |
| r over Q /$\Omega$ | 147.84 |
| intrinsic quality factor $Q_0$ | $8\times10^8$ |
| optimum coupling $\beta_{opt}$ | 800 |
| transit time factor $T_{tf}$ | 0.8 |
| forward generator power $P_g$ /kW | 6.7 |
| phase of the generator current $\theta$ | $-\pi/6$ |
| synchronous phase $\phi$ | $-\pi/6$ |
| half phase width of the bunch $\Delta\phi_{0.5}$ | $\pi/12$ |
| design beam current $I_b$ /mA | 10 |

For the superconducting cavity, the optimum detuning angle of the cavity under the designed beam current intensity can be well approximated as[6, 8] $\psi_{opt} \approx -\phi = \pi/6$.

Due to the space limitations, it's impossible to list all of these evaluations here. Instead, the evaluation for only one typical transient beam loading process will be demonstrated in this section. The process for the evaluation are briefly described as below.

At $t_0 = 0$, the generator is turned on and cavity field buildup process begins. At $t_1 = \Delta T$, the field in the cavity reaches to the steady state I, when the CW beam injection begins. Due to the beam loading effect, the field in the cavity will evolve to the steady state II. At $t_2 = 2\Delta T$, the generator power suddenly drops to zero and beam continues passing through the cavity, resulting to the steady state III of the cavity field. At $t_3 = 3\Delta T$, the generator power is recovered, the field in the cavity will restore to the steady state II.

Obviously, $\Delta T$ should be sufficiently larger than the cavity time constant $\tau$ to reach the steady state, therefore we let $\Delta T = 8\tau$ in the following evaluation.

The time evolution of the generator power and beam current are shown as below.

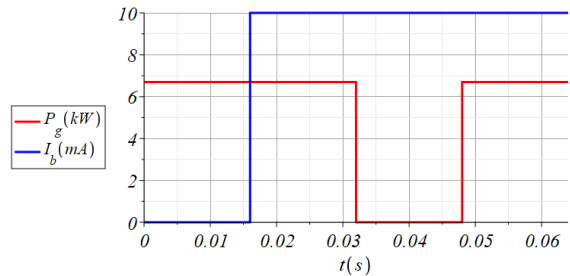

Fig. 2 The time evolution of the generator power and beam current

The trace of cavity voltage on the complex plane and time evolution of the reflected power from the cavity had been calculated with (3.1.3) and (3.1.1) respectively for two cases of the cavity detuning, including the optimum detuning and the extreme detuning. The obtained results are shown as below,

(1) $\psi = \psi_{opt}$

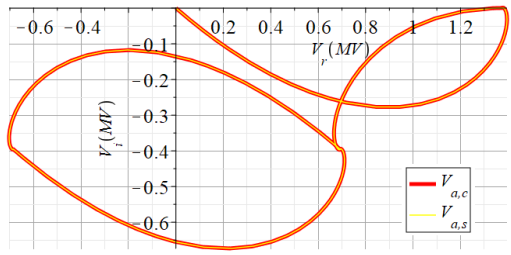
(a) Trace of the Cavity Voltage on the Complex Plane

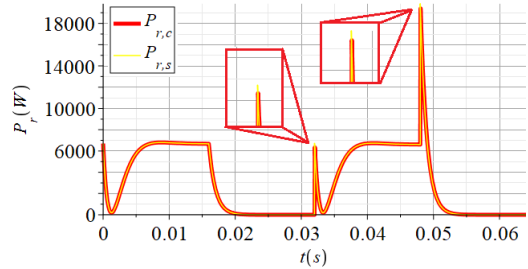
(b) Reflected Power of the Cavity

(2) $\psi = 9\pi/20$

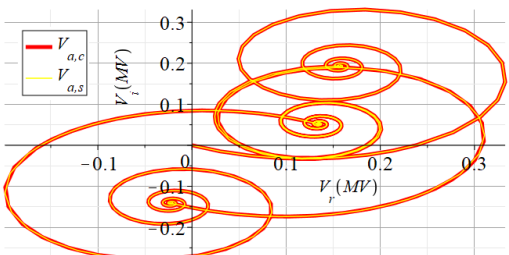
(c) Trace of the Cavity Voltage on the Complex Plane

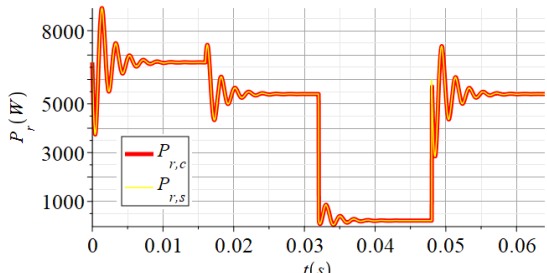
(d) Reflected Power of the Cavity

Fig. 3 The trace of cavity voltage and reflected power from the continuous and step function
(a) and (b) are for the optimum detuning, while (c) and (d) are for the extreme detuning

A technique was adopted to plot Fig. 3, the thinner yellow curve is plotted above the thicker red curve, which means if the deference between them is small enough, the yellow curve should be enclosed by the red one, i.e., the compound curve consisting of the red and the yellow curve should have a closed red boundary.

By observing Fig. 3, it can be seen that curves in Fig. 3 (a), (c) and (d) all have got closed red boundaries. As for the curve in Fig. 3 (b), the red boundary is open at the two sharp peaks. If we look closer at these two locations, we'll find out that the relative deference between the red and yellow curve are still very small. For the first peak in Fig. 3 (b), the relative difference is $5.26\%$, while $2.60\%$ for the second peak. The two curves at the first peak in Fig. 3 (b) is plotted as the following to facilitate the comparison.

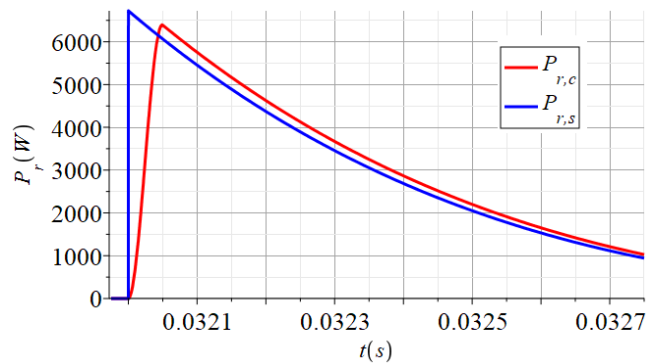

Fig. 4 Two curves of reflected power at the first peak

Various situations have been evaluated, finding that the difference in cavity voltage and reflected power under the two forms of the effective driven current are always relatively small. For most cases, these differences are even smaller than the difference between the equivalent circuit model and the reality.

Furthermore, evaluations with a broad range of superconducting cavity and beam parameters shows their agreements with the above conclusion, which means the step function approximation of the effective driven current is good enough for the transient beam loading analysis for the superconducting cavity.

## 4. Comparison Between the Analytical and Numerical Results

The transient beam loading process in the previous section can also be numerically solved. With (3.1.3), the numerical results obtained by the Fehlberg fourth-fifth order Runge-Kutta method are plotted in Fig. 5.

(1) $\psi = \psi_{opt}$

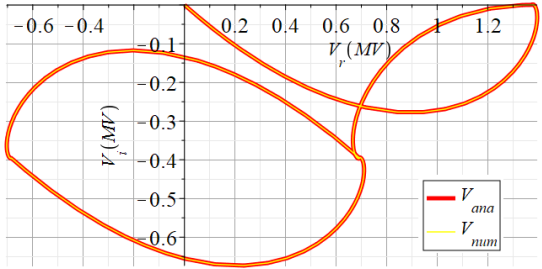
(a) Trace of the Cavity Voltage on the Complex Plane

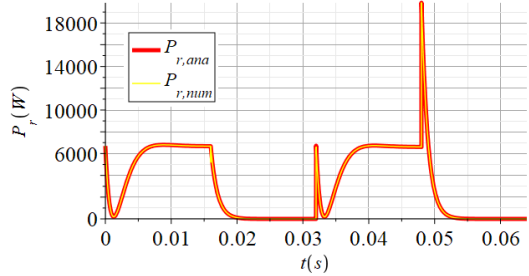
(b) Reflected Power from the Cavity

(2) $\psi = 9\pi/20$

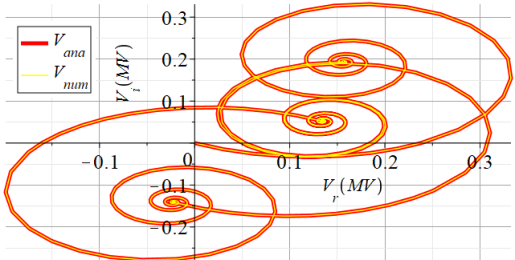
(a) Trace of the Cavity Voltage on the Complex Plane

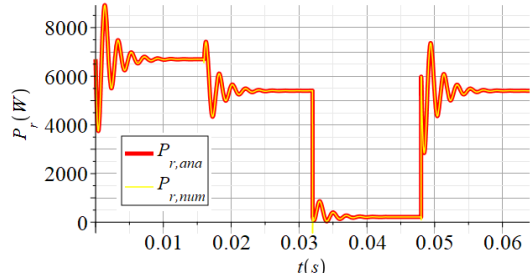
(d) Reflected Power from the Cavity

Fig. 5 The trace of cavity voltage and reflected power from the analytical and numerical methods
(a)and (b) are for the optimum detuning, while (c) and (d) are for the extreme detuning

The same technique adopted in Fig. 3 was also used in Fig. 5 to facilitate the comparison. It can be seen all the curves in Fig. 5 have got closed red boundaries, indicating the difference between the analytical and numerical results are small enough. To obtained numerical results with such accuracy, the computation time is always larger than that of the analytical method (Excluding the calculation time for the analytical solution, which is once for all). And the difference in computation time for the two methods will increases with $|\psi|$.

In general, because (2.1.24) is linear, there is no butterfly effect. The accuracy of the numerical results can always be guaranteed by sufficient small step-size in the solving process, at the cost of increasing the computation time. And as $\psi \to \pm \pi/2$, the computation time for numerical solving will tend to infinity.

## 5. Conclusion

The driven term in the dynamic equation for the superconducting cavity voltage can be well approximated into a step function and the transient beam loading problem of the superconducting cavity can be analytically solved. For the analytical solution, besides its accuracy advantage, great amount of time can be saved for evaluations under various sets of parameters to investigate the transient beam loading effect for different situations. For the numerical solving method, each set of parameters must be run separately to obtained the time evolutions of the cavity voltage, reflected power from the cavity and other interested quantities. While for the analytical solving method, the solving process can be done once for all. Just by substituting different sets of parameters into the analytical solution, the corresponding time evolutions of the interested quantities can be obtained instantly.